\documentclass[amsfonts, amssymb, amsmath, aps, 10pt, pra, twocolumn, superscriptaddress, english, floatfix]{revtex4-2} 
\usepackage[utf8]{inputenc}
\usepackage[english]{babel}
\usepackage[T1]{fontenc}
\usepackage{mathtools}
\usepackage{physics} 
\usepackage{graphicx}
\usepackage[dvipsnames]{xcolor}
\usepackage[colorlinks=true, linkcolor=RoyalBlue, urlcolor=RoyalBlue, citecolor=RoyalBlue]{hyperref}

\usepackage{graphicx}
\graphicspath{{./figures_PDF_of_F/}}

\usepackage{multirow}
\usepackage{booktabs} 
\usepackage{array}
\newcolumntype{L}{>{$}l<{$}}
\newcolumntype{C}{>{$}c<{$}}
\newcolumntype{P}[1]{>{\centering\arraybackslash}p{#1}}
\newcolumntype{M}[1]{>{\centering\arraybackslash}m{#1}}

\usepackage{nameref}
\usepackage[nameinlink]{cleveref}
\Crefname{appsec}{Appendix}{Appendices}
\Crefname{box}{Box}{Box}
\Crefname{equation}{Eq.}{Eqs.}
\Crefname{figure}{Fig.}{Figs.}

\newcommand{\calH}{\mathcal{H}}

\newcommand{\lavg}{\langle}
\newcommand{\ravg}{\rangle}
\newcommand{\calA}{\mathcal{A}}
\newcommand{\calAb}{{\mathcal{A}^{\mathsf{c}}}}
\newcommand{\calB}{\mathcal{B}}
\newcommand{\calBb}{{\mathcal{B}^{\mathsf{c}}}}
\renewcommand{\Tr}[2]{\text{Tr}_{#1}{#2}}

\newcommand{\pdf}{\text{pdf}}

\begin{document}

	\title{Distribution of Fidelity in Quantum State Transfer Protocols}
	
	\author{Salvatore Lorenzo}
	\affiliation{Universit\`a degli Studi di Palermo, Dipartimento di Fisica e Chimica - Emilio Segr\`e, via Archirafi 36, I-90123 Palermo, Italy}
	\author{Francesco Plastina}
	\affiliation{Dipartimento di Fisica, Universit\`a della Calabria, 87036 Arcavacata di Rende (CS), Italy} 
	\affiliation{INFN, gruppo collegato di Cosenza}
	\author{Tony J. G. Apollaro}
	\affiliation{Department of Physics, University of Malta, Msida MSD 2080, Malta}
	\author{Mirko Consiglio} 
	\affiliation{Department of Physics, University of Malta, Msida MSD 2080, Malta}
	\author{Karol \.{Z}yczkowski}
	\affiliation{Institute of Theoretical Physics, Jagiellonian University,  Łojasiewicza 11, 30-348 Krak{\'o}w, Poland}
	\affiliation{Center for Theoretical Physics, Polish Academy of Sciences, Al. Lotnik\'{o}w 32/46, 02-668 Warszawa, Poland}
	
	\date{April 29, 2024}
	
	\begin{abstract}
Quantum state transfer protocols are a major toolkit in many quantum information processing tasks, from quantum key distribution to quantum computation. 
To assess performance of a such a protocol, one often relies on the average fidelity
between the input and the output states. Going beyond this scheme, we analyze the entire
probability distribution of fidelity, providing a general framework to derive it for 
the transfer of single- and two-qubit states. 
Starting from the delta-like shape of the fidelity distribution, characteristic to 
perfect transfer, we analyze its broadening and deformation due to
realistic features of the process, including  non-perfect read-out timing.
Different models of quantum transfer, sharing the same value of the average fidelity, display different distributions of fidelity, providing thus additional information on the protocol, 
including the minimum fidelity.
	\end{abstract}
	
	\maketitle
	\section{Introduction}
	
	Faithful transfer of quantum information between different
	locations or different parts of the same quantum hardware is a  basic building block of several quantum information processing protocols and it is crucial for a variety of applications~\cite{kimble2008a, cuomo2020a}. 
 Since the seminal paper by Bose~\cite{bose2003a}, spin chains emerged as suitable models to describe quantum data buses allowing one for high fidelity quantum state transfer (QST). One usually considers
 short distance communication,
 or more generally, the transfer of a state
  from site ${\cal A}$ to ${\cal B}$,
  typically located at two ends of the chain.
	
	A great variety of different protocols which employ spin systems as transmission channels have been devised and experimentally realized~\cite{vogell2017,li2018,ali2018,chapman2016,xiang2024enhanced}, achieving either `perfect' or `almost perfect' QST~\cite{vinet2012}. Broadly speaking, they can be divided into different classes, according to the physical mechanism on which they rely in order to achieve the transmission -- see e.g. Ref.~\cite{apollaro2013a}. Examples consist of protocols based on time-dependent couplings~\cite{difranco2010a}, including pulsed and general non-adiabatic driving~\cite{agundez2017a,huang2018a,wang2020b,kiely2021a,ji2022a}, fully engineered interactions~\cite{christandl2004a,difranco2008a,markiewicz2009,karimipour2012,serra2022a}, ballistic transfer~\cite{banchi2010a,banchi2011a,apollaro2012a}, Rabi-like oscillations between eigenstates having support on the sending and receiving sites~\cite{PhysRevA.72.034303, gualdi2008a,linneweber2012,lorenzo2013c, chen2014a, korzekwa2014a}. 
 All of these methods have been shown to achieve transmission perfectly or with a high fidelity. These approaches have been generalized in various ways  to describe setup
 with more than one receiver,
 to study routing of quantum states~\cite{paganelli2013a, yousefjani2020a}, distribution of entanglement~\cite{almeida2018,apollaro2020a,vieira2020,wang2020b,noi23}, shared between two 
 or among several qubits~\cite{lorenzo2015c,apollaro2015a,chetcuti2020a,apollaro2022b}.
 State transfer in the presence of noise and disorder due to fluctuating local fields along the spin chain analyzed in  ~\cite{pavlis2016,almeida2018,keele2022a}, 
	has been simulated with a quantum computing device~\cite{babukhin2022a}.
	
	To  assess the quality of the state transmission through the spin chain quantitatively, it is customary to employ the average fidelity. Indeed, the fidelity, $F \in [0,1]$, provides a measure of the overlap between the input state of the sending part, $\ket{\psi_{\cal A}}$, with the output state
 $\rho$ of the receiving part. Its expectation value, $\langle F \rangle = \langle \psi_{\cal A}|\rho|\psi_{\cal A}\rangle$, averaged  over the set of all pure states of the sender, is an indicator of the average quality of the transmission protocol.
	
	However, in several cases, the average fidelity does not provide a full characterization of the transfer process, which can, instead, be obtained by analysing the full statistics of the values taken by the fidelity over the Hilbert space of the sending part. This information is contained in the probability distribution function (PDF) of fidelity.
	
	The only case in which the distribution of the fidelity does not provide any new information with respect to the average is that of perfect state transfer, where the distribution becomes a delta function, peaked on the value $F=1$. However, a perfect QST typically poses severe requirements, such as an extremely high level of control of the parameters of spin network, or a very precise extraction time at the receiving site. If these requirements are not met, the ideal PDF of the fidelity can acquire a width, and its functional form may provide us with a way to characterize the physical process underlying the state transfer.

     Distribution of fidelity between random quantum states was analyzed in \cite{ZS05},
     while ref. \cite{CRZ23} studies the distribution between random input pure states
     and the output an a given quantum operation. 
      The aim of this work is to investigate the
      distribution of fidelity    
       for state transfer in  
	the case of a non-precise read-out timing on the receiving side.
	This assumption leads to two specific forms of the distribution depending on the transmission mechanism. Thus, we will demonstrate explicitly that the equivalence at the level of the average fidelity is not enough to conclude that different protocols are fully equivalent, as they can give rise to different distributions of fidelity
    in presence of imperfections.

    This paper is organized as follows.
    In \Cref{probabdistr}  
    we highlight techniques to obtain the distribution of  fidelity for 
    state transfer protocols. After recalling
     in \Cref{transfermap}
    a general approach to evaluate the fidelity  for spin-chain-based transfer using the Kraus representation of the channel connecting the sender to the receiver, we apply the general method in \Cref{results}, in which
	three different protocols are analyzed. 
 We consider the one-qubit and two-qubits
   state transfer protocol. In the latter case the distribution fidelity is derived for states with a fixed value of entanglement. The paper is concluded in \Cref{conclusions} with some final remarks.


 \section{Distribution of fidelity}
 \label{probabdistr}
 For a given protocol, we can evaluate the fidelity for any sender state; its expression will vary with the details of the dynamics, but, generally speaking, it will depend on the parameters with which we identify the state to be sent, and on the time elapsed since the beginning of the protocol,  i.e. $F\equiv F(t;\alpha,\beta,\dots)$.  To obtain the probability distribution of fidelity
 at a given time $t$
 we need to exploit its functional dependence on the state parameters.
	
  Denote the mean  fidelity
  averaged over the others parameters
  $F(\alpha)=\int F(\alpha,\beta,\cdots)d\beta\cdots$.
  With a minor abuse of notation, 
  we can write the inverse functions,
   $\{\alpha (F),\beta(F),\cdots\}$.  
  Then  we arrive at the following form of the
    probability distribution function,
	\begin{equation}
 \label{pdftransf}
		\pdf(F)=\pdf \left(\alpha(F)\right) \left|\frac{\partial\alpha(F)}{\partial F} \right|
		, \;\;\;\;\; \text{with}\;\;0\leq F\leq 1.
	\end{equation}

The primary concern at this point is to derive a clear expression for the fidelity over time, and to establish a parametrization that allows us to utilize \Cref{pdftransf}. Generally, this task is not straightforward. Therefore, in the subsequent discussion, we aim to outline a road-map to achieve the desired outcome.

	\section{General approach to the transfer map} \label{transfermap}
	We consider a general network system in which each node hosts a $d$-level system, with Hilbert space given by $\calH=\mathbb{C}^d$.	
	As pointed out above, we are interested in evaluating the fidelity between a pure input state $\ket{\psi_{\cal A}}$ written at $t=0$  on a set of sending sites $\calA$, and the state $\rho_\calB(t)$ retrieved, at the end-time of the protocol, $t$, from a different set of receiving sites $\calB$; namely,
	\begin{equation}\label{fid}
		F(t)=\bra{\psi_\calA}\rho
		_\calB(t)\ket{\psi_\calA}.
	\end{equation}
	
	Assuming that the rest of the system, which we call $\calAb$, is initialized in \emph{fixed} state $\ket{\psi_\calAb}$, the total initial state is $\ket{\psi_\calA}\otimes\ket{\psi_\calAb}$ and the reduced density matrix of the receiving sites $\calB$ is obtained after tracing out the complementary degrees of freedom, $\calBb$:
	\begin{align}
		\label{rhoB}
		&\rho_{\calB}(t)
		{=}\Tr{\calBb}{\left\{U(t)\Big(\!\ket{\psi_\calA}\otimes\ket{\psi_\calAb}\!\Big)\Big(\!\bra{\psi_\calA}\otimes\bra{\psi_\calAb}\!\Big)U^\dagger(t)\right\}}\nonumber \\
		&{=}\sum_{\phi_\calBb}\bra{\phi_\calBb} U(t) \ket{\psi_\calAb} \!\Big(\ket{\psi_\calA}\bra{\psi_\calA}\Big)\!\bra{\psi_\calAb}U^\dagger(t)
		\ket{\phi_\calBb}~,
	\end{align}
	\begin{figure}[t]\centering
		\includegraphics[width=0.8\linewidth]{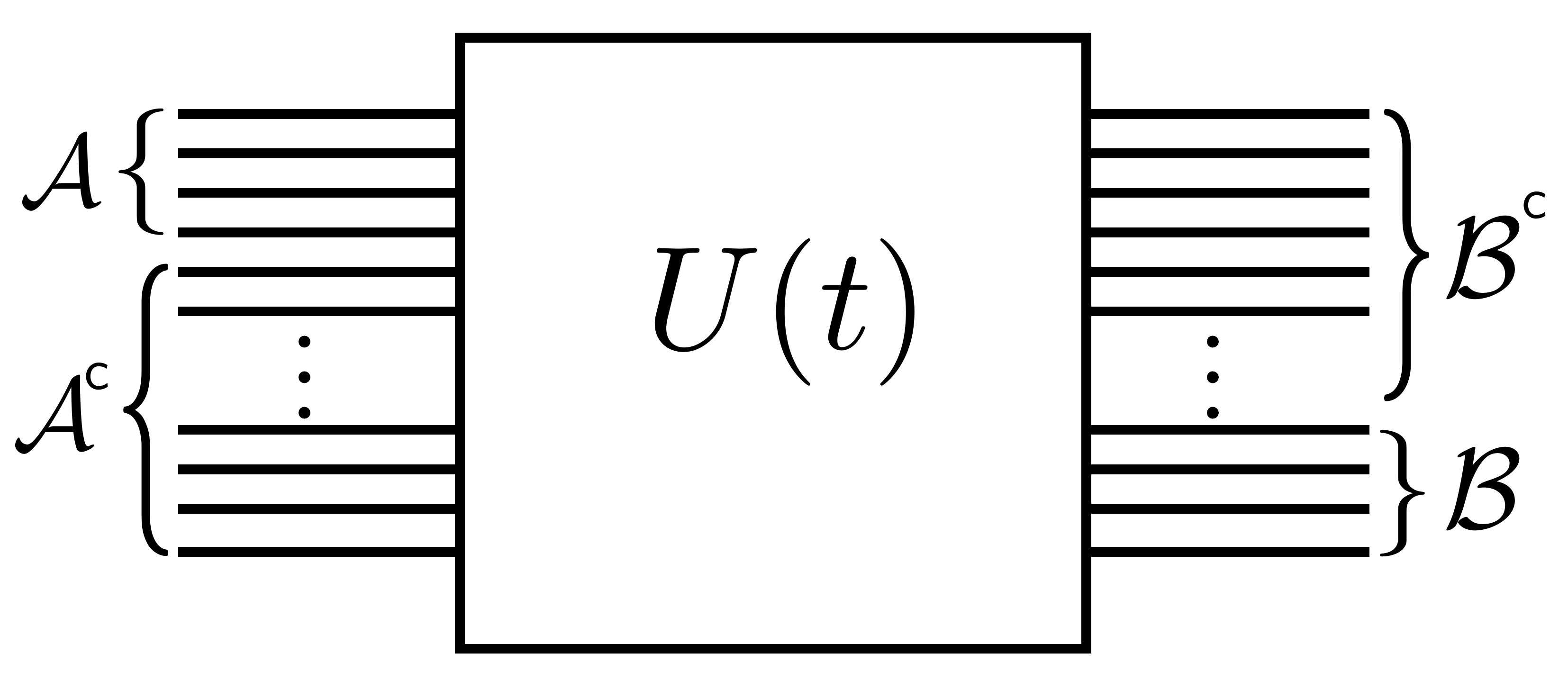}
		\caption{Sketch of the transmission and illustration of our notation: the state $\ket{\psi}$ is prepared at the sending sites, ${\cal A}$, and its information content is spread all over the system, ${\cal A} \bigcup {\cal A}^c$, due to the unitary evolution $U$. The state is later retrieved at the receiving sites ${\cal B}$
			.}\label{fig1}
	\end{figure} 
	\noindent where $U(t)$ is the unitary time evolution operator describing the dynamics of the full system, and $\{ \ket{\phi_{\calBb}}\}$ denotes a complete orthonormal basis of $\calBb$. The second line of \Cref{rhoB} provides the Kraus decomposition of the completely positive linear operator $\Lambda(t):\calH_\calA{\rightarrow}\calH_\calB$, which maps the initial state of $\calA$ onto the state of $\calB$ at time $t$. Explicitly,
	\begin{equation} \label{map}
		\rho_\calB(t){=}\Lambda(t)\Big[\ket{\psi_\calA}\bra{\psi_\calA}\Big]{=}\sum_{\phi_\calBb}E_{\phi_\calBb}\ket{\psi_\calA}\bra{\psi_\calA}E_{\phi_\calBb}^\dagger,
	\end{equation}
	with $E_{\phi_\calBb}(t)= \bra{\phi_\calBb} U(t) \ket{\psi_\calAb}$. Substituting this expression for $\rho_\calB$ into \Cref{fid}, we obtain the fidelity in the form 
	\begin{equation}\label{fid2}
		F(t)=\sum_{\phi_\calBb}\vert\bra{\psi_\calA}E_{\phi_\calBb}(t)\ket{\psi_\calA}\vert^2.
	\end{equation}
	This general expression for $F$ will be utilised in the following for different state transfer channels.
	
	\subsection{Dynamics with U(1) symmetry}
	The Kraus operators $E_{\phi_\calBb}(t)$ entering \Cref{fid2} can be cumbersome to evaluate for arbitrary channels. However, in the presence of symmetries in the system's dynamics, some general assumptions can be made about their form. Let us assume that a conserved quantity $\mathbf{Q}$ exists whose spectral decomposition (taking into account possible degeneracies) is given by
	\begin{equation}
		\mathbf{Q}=\sum_q q \sum_{d=1}^{d_q}\ket{q_d}\bra{q_{d}}~.
	\end{equation}
	Then the evolution operator $U(t)$ reads
	\begin{equation}\label{UinQ}
		U(t)=\sum_{q}\sum_{d=1}^{d_q}\sum_{d'=1}^{d_q}U_{q}^{dd'}(t)\ket{q_d}\bra{q_{d'}}=U_1\oplus U_2\oplus\cdots, 
	\end{equation}
	where $U_q^{dd'}=\bra{q_d}U(t)\ket{q_{d'}}$. 
   The above equation entails 
	that $U(t)$ does not couple different subspaces of the conserved quantity $\mathbf{Q}$.
	In order to take advantage of the presence of invariant subspaces, each Kraus operator appearing in \Cref{map} can be expressed as
	\begin{align}\label{kraus}
		E_{\phi_\calBb}{=}&\bra{\phi_\calBb} U(t)\ket{\psi_{\calAb}}
		= \nonumber\\ 
		=& \sum_{\phi_\calB}\sum_{\phi_\calA}\bra{\phi_\calB}\bra{\phi_\calBb}U(t)\ket{\psi_{\calAb}}\ket{\phi_{\calA}}\;\ket{\phi_\calB}\bra{\phi_{\calA}}.
	\end{align}
	Inserting \Cref{UinQ} into the latter expression, we realize that if neither $\ket{\phi_\calB,\phi_\calBb}$ nor $\ket{\psi_{\calAb},\phi_{\calA}}$ belong to the $q$-subspace, the corresponding element of $E_{\phi_\calBb}$ is null.
	
	If we now assume that  $\mathbf{Q}$ is a sum of {\it local} observables, namely
	\begin{equation}
		\mathbf{Q}{=}\sum_{i=1}^{N}\mathbf{q}_i{=}\!
		\left(\sum_{i\in \calA}\mathbf{q}_i{+}\!\sum_{j\in\calAb}\mathbf{q}_j\right)\!{=}\!
		\left(\sum_{i\in \calB}\mathbf{q}_i{+}\!\sum
		_{j\in\calBb}\mathbf{q}_j\right)~,
	\end{equation} 
	where $\mathbf{q}_i$ is a \emph{local} operator acting on each local $\calH_i$, then $q=q_\calAb+q_\calA=q_\calBb+q_\calB$, which implies that only the Kraus matrix elements fulfilling the above conditions are non-zero.
	In particular, these considerations apply to the case of a spin system with, e.g., conserved magnetization along the quantization axis, holding for Hamiltonians with global $\mathrm{U}(1)$ symmetry. 
	
	\section{Spin chain-based state transfer}
	\label{results}
	In this section we provide some explicit cases in order to apply the general framework outlined above. In particular, we will consider  the transfer of one and two spin-$1/2$ states over a network of coupled spins. In this case, we take the sender to reside at the first sites of the chain, while the receiver is located at the opposite edge, e.g., at site $N$ for the single qubit state transfer, or $\{N{-}1,N\}$ in the $2$ qubit case.
	
	We consider a network $\mathbf{G}$ of $N$ sites, each hosting a qubit, 
 whose dynamics is generated by an $\mathrm{U}(1)$-symmetric Hamiltonian of the form:
	\begin{equation}\label{BoseH}
		\mathbf{H}_{\mathbf{G}}=\sum_{i\neq j}^{N} J_{i j}\left(\sigma^i_x\sigma^j_x{+}\sigma^i_y\sigma^j_y{+}\Delta_{i j}\sigma^i_z\sigma^j_z\right)+\sum_{i=1}^N B_i \sigma_z^i~,
	\end{equation}
	where $\sigma_{x / y / z}^i$ are the Pauli matrices for the qubit residing at the $i$-th site, $B_i$ are static magnetic fields and $J_{i j},\Delta_{i j}$ are coupling strengths.
	
	This Hamiltonian is symmetric under rotation around the $z$-axis; thus, in this context, the quantity $\mathbf{Q}=\sum_{i=1}^N \sigma_i^z$ is, indeed, conserved. 
	\subsection{One-qubit transfer}
	The first case under investigation is the transfer of the state of one qubit.
	This entails that the only allowed values of $q$ are $0+q_\calAb$ and $1+q_\calAb$, where $q_\calAb=\expval{\sum_{j\in\calAb}\sigma^j_z}{\psi_\calAb}$. Hence, there are only two classes of
	Kraus operators corresponding to $q_\calBb=0,1$ and their explicit expressions depend on the state of the channel at the time the transfer protocol begins, which we specify in the following subsections.
	\paragraph*{Initial vacuum state for the channel ---}\label{1qubit_vacuum}
	If we take $\ket{\psi_\calAb}=\ket{0\dots 0}$, then $q_\calAb=0$ and the Kraus operators in \Cref{kraus} are given by
	\begin{equation}\label{kraus1}
		E_0{=}
		\begin{pmatrix}1 & 0\\0 & a_1^N(t)\end{pmatrix},
		\;
		E_1^{j}{=}
		\begin{pmatrix}0 & a_j^N(t)\\0 & 0\end{pmatrix}\;\text{for}\;j{=}2,\hdots,N,
	\end{equation}
	where we have renamed $U_1^{ij}(t)=a_i^j(t)$.
	The class $q_\calBb=1$ can be cast as a single Kraus operator,
	\begin{align}
		E_1{=}
		\begin{pmatrix}0 & \sqrt{1-\left|a_1^N(t)\right|^2}\\0 & 0\end{pmatrix}~.
	\end{align}
	By parametrizing the sender state as 
	\begin{align}
		\label{eq_state_1q}
		\ket{\psi_\calA}=\cos\left(\frac{\theta }{2}\right)\ket{0}+e^{i\phi}\sin\left(\frac{\theta }{2}\right)\ket{1}, \nonumber \\
		\text{with}\;\; 0\leq\theta\leq\pi\; \;\text{and}\; \; 0\leq\phi\leq 2\pi,
	\end{align}
	the fidelity~\eqref{fid2} turns into a function of $\theta$, without $\phi$ dependence, due to the fact that the Hamiltonian in \Cref{BoseH} possesses $\mathrm{U}(1)$ symmetry and the initial state of the channel $\ket{\psi_\calAb}$ is an eigenstate of the total magnetization $\mathbf{Q}$.
	It turns out that
	\begin{equation}\label{fid3}
		F(t,\theta)=
		\bigl(1-\vert a_1^N(t)\vert^2\bigr) s^2 c^2+
		\left\vert a_1^N(t) s^2+c^2\right\vert^2,
	\end{equation}
	where we used the completeness relation $\sum_{j=1}^N \vert a_j^N\vert^2=1$ and defined $s\equiv\sin(\theta/2), c\equiv\cos(\theta/2)$. 
	
	The PDF for the angle $\theta$ of uniformly distributed pure states on the Bloch sphere is given by
	\begin{equation}\label{pdftheta}
		\pdf(\theta)=\dfrac{1}{2}\sin \theta, \qquad \text{with}\;\;0\leq\theta\leq\pi,
	\end{equation}
	so that, by straightforward application of \Cref{pdftransf}, we obtain a closed expression for the distribution of fidelity  reported in \Cref{appA}.
	
\begin{figure}[t]
		\centering
		\includegraphics[width=0.95\linewidth]{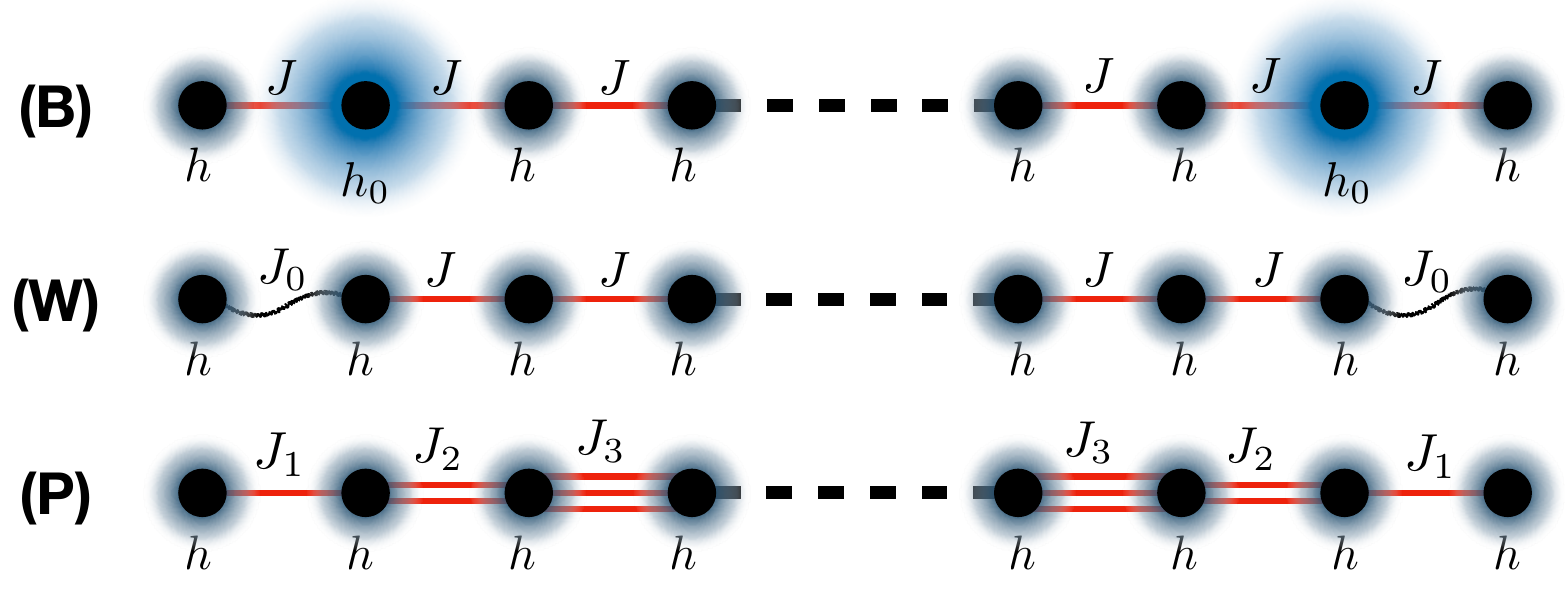}
		\caption{Sketch of the three protocols considered in this work. The  coupling constants $J_n$ are depicted by red segments, whose number represent the strength of coupling. In the W protocol, the wavy segment stands for a very weak $J$. The dots symbolize the sites, while their shadows represent the local magnetic fields, where a larger shadow indicates a stronger magnetic field. The barrier (${\mathbf B}$) protocol assumes uniform couplings and a strong magnetic field applied to the nearest-neighbor of the senders and the receivers. The weak coupling (${\mathbf W}$) protocol is defined by weaker couplings between the chain and the senders (receivers). In the perfect (${\mathbf P}$) protocol one assumes to be able to engineer all of the couplings between the sites. Details of each protocol are discussed in \Cref{appB}.}
		\label{chains_sketch}
	\end{figure}

\begin{figure}[t]\centering
		\includegraphics[width=0.9\linewidth]{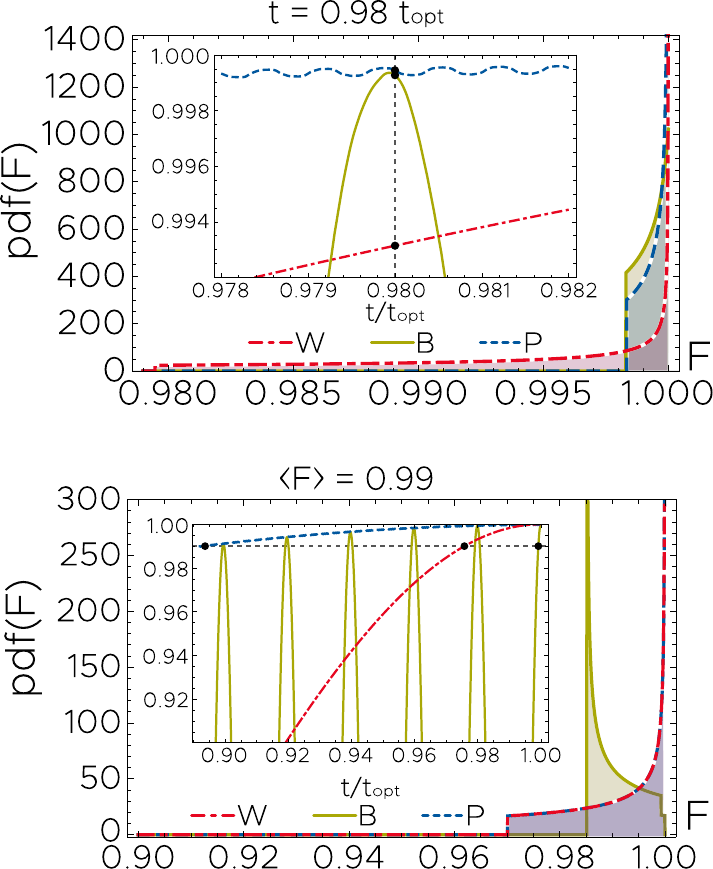}
		\caption{ Probability distribution of fidelity for three protocols analyzed: (top) Reading time is set to up the $2\%$ of the optimal one, which depends on the choice of the protocol. The average fidelity, in all of the three cases, is greater than $0.99$ -- see the insets;  (bottom) Average fidelity  is fixed to $0.99$ and the minimal  fidelity of the protocol B is higher when compared to the W- and P-protocols. Results are obtained for  $N=22$ sites, with $h_0=200$ for the B-protocol, and $J_0=1/200$ for the W-protocol -- see \Cref{appB} for details. }\label{pdf_1Qubit}
	\end{figure}

	
	We are now in position to compare the distribution of fidelity obtained from different transfer protocols. We will consider the weak-coupling approach (W) \cite{PhysRevA.72.034303}, the barrier protocol (B) \cite{lorenzo2013c}, and the perfect transfer protocol (P) \cite{christandl2004a} -- see \Cref{chains_sketch} and \Cref{appB} for details. A sketch of these three protocols  provided in \Cref{chains_sketch} emphasizes their main differences: the coupling constants $J_n$ are depicted by red segments, whose number indicates the strength of coupling. For the W protocol, the wavy segment stands for a very weak $J$. Shadows around the dots are used to represent local magnetic fields: the larger the shadow, the stronger the field.
	All of these approaches admit an optimal time transfer $t_\text{opt}$ at which the average fidelity over all possible input states reaches the maximum value. It is exactly $\lavg F \ravg =1$ for the P-protocol, whereas for the W- and B-protocols the unit value is reached asymptotically, although at the expense of a longer transfer time.
	This means that, for all practical purposes, at the optimal time the protocol is perfect, so that
     $\pdf (F)=\delta(F-1)$. 
	On the other hand, if the read-out timing is not precise, or if, for experimental reasons, one is satisfied with an $\lavg F \ravg$ slightly smaller than $1$, by choosing a ``compromise setting'',
	then, different protocols display very different behaviours of the $\pdf$, as reported in \Cref{pdf_1Qubit}.
	
	In particular, as one can notice from the insets in \Cref{pdf_1Qubit}, and discussion in \Cref{appB}, these three protocols show different sensitivities to the optimal time. The average fidelity for the B-protocol, oscillates in time and shows very narrow peaks, quite close to each other. On the other hand, the P-protocol and the W-protocol display a single, but much broader peak, with a reduced dependence on the choice of the time $t_\text{opt}$.
	
	As a result of the different dynamics, the W- and B-protocols display quite a similar distribution of fidelities, with a lower bounded support. The $\pdf(F)$ for the P-protocol, instead, shows tails extending to much smaller values of $F$, as seen from the top panel of \Cref{pdf_1Qubit}.
	In the lower panel, we show the fidelity distributions under the condition that the average fidelity is set to a smaller value, $\lavg F \ravg < 1$, common to all of the three protocols. In this case, one can see that the W- and P-protocols, display the same $\pdf(F)$, peaked at $F=1$ and with a tail extending down to a minimum fidelity value, $F_{min}$. The $\pdf(F)$ for the B-protocol, instead, has a peak at a smaller $F$, with a tail extending on its right side, up to $F=1$.
	
	The features displayed by the PDFs can be understood by analyzing the dynamics of the spin system in the three cases. In particular, setting the average fidelity to a given value amounts to choosing both the absolute value of the transition amplitude $a_1^N$ and its real part (see the expression for $\lavg F \ravg$ reported in \Cref{appB}). 
	It turns out that for both the P- and W-protocols, this amplitude can be taken to be either real or purely imaginary, so that it is fully determined by the choice of $\lavg F \ravg$. Thus, these two approaches to state transfer have the same amplitude for a given $\lavg F \ravg$. Now, by virtue of \Cref{eq_fid_1}, they enjoy the same distribution with $\theta$.
	
	Interestingly, from \Cref{fid3}, it is also possible to derive the states that are transmitted with minimum fidelity at fixed transition amplitude $a_1^N$. This is an important quantity in QST protocols as it allows the definition of a lower bound on the reliability of the transferred state. For the P- and W-protocols, where the transition amplitude $a_1^N$ is real (for $N$ odd) or imaginary (for $N$ even), we can set $a_1^N$ to be real (possibly by application of a constant magnetic field in the even $N$ case). Hence the state with $\theta=\pi$, i.e., $  \ket{\psi_\calA}=\ket{1}$, yields the minimum fidelity $F(t,\pi)=\left(a_1^N\right)^2=\left(\sqrt{2\left(3 \lavg F \ravg-1\right)}-1\right)^2$ for all values of $a_1^N$. 
	On the other hand, for the B-protocol, where no constant magnetic field is applied for maximising the average fidelity, the state that minimizes \Cref{fid3} is dependent on the complex amplitude $a_1^N$.
	Expressing the transition amplitude in polar form $a_1^N=r e^{i \phi}$, we obtain that the minimum fidelity obtained in \Cref{fid3} for
	\begin{align}
		\label{eq_min_fid}
		\begin{cases}
			\theta=
			\cos^{-1}\left(\frac{r^2-1}{2r\left(r-\cos\phi\right)}\right), &r>\frac{1}{3}~\text{and}~\phi \in \left[\phi^*,
			2\pi-\phi^*\right]\\
			\theta=\pi, &r>\frac{1}{3}~\text{and}~\phi \notin \left[\phi^*,
			2\pi-\phi^*\right]
			\\
			\theta=\pi, &r\leq\frac{1}{3}~\text{and}~0\leq\phi< 2\pi
		\end{cases}~
	\end{align}
	where $\phi^*=\cos^{-1}\left(\frac{3r^2-1}{2r}\right)$.
	\Cref{eq_min_fid} illustrates the fact that, for the cases of interest where $r$ approaches unity, the state that mimimizes the fidelity in \Cref{fid3} can be either $\ket{\psi_\calA}=\ket{1}$ or a state lying in the neighborhood of the equatorial plane as $\theta \rightarrow \pi/2$, depending on the phase $\phi$ of the transition amplitude $a_1^N$.
	Equipped with the knowledge of the state that achieves the maximum fidelity, i.e., $\ket{0}$ for all protocols, and the state that achieves the minimum fidelity we can now explain the positions of the peaks in the $\pdf(F)$ in \Cref{pdf_1Qubit}. For the W- and P-protocol, the rate of change of 
	the fidelity with $\theta$ is less at $\theta=0$ than at $\theta=\pi$, implying that there are more states in the neighborhood of $\ket{0}$ than in that of $\ket{1}$ in a fixed interval of the maximum and minimum fidelity, respectively. On the other hand, for the B-protocol, when the minimum fidelity is achieved for a state lying close to the equatorial plane, the opposite holds.
	\paragraph*{$\ket{\psi_\calAb} \neq \text{vacuum}$ ---}In order to show the role played by the Kraus operators, now we assume that the initial state $\ket{\psi_\calAb}$ belongs to the 1-excitation subspace and that it is uniformly spread over the entire system:
	\begin{equation}
		\label{not_vacuum}
		\ket{\psi_\calAb}=\frac{1}{\sqrt{N{-}2}}\sum_{j=2}^{N-1}\ket{j}.
	\end{equation}
	\begin{figure}[t]\centering
		\includegraphics[width=0.9\linewidth]{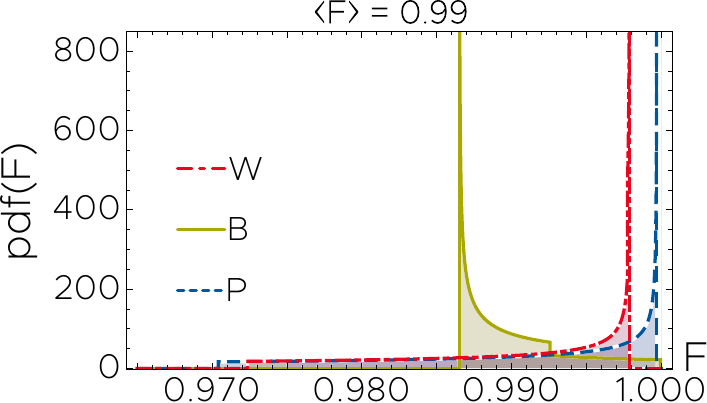}
		\caption{\label{1_qubit_not_vacuum} PDF of the fidelity for $\ket{\psi_\calAb}$ when the channel is in the single-excitation state given by \Cref{not_vacuum}. The three protocols have very different behaviour despite their average fidelity being fixed to $\lavg F\ravg=0.99$. The $B$-protocol shows a higher minimum fidelity, compared to $W$ and $P$ protocols. The latter protocols, however, transfer more states with higher fidelity. The fugure reports the case for $N=15$, $h_0=100$ for the B protocol and $J_0=1/100$ for the W protocol.}
	\end{figure}
	With these assumptions we have three different classes of Kraus operators corresponding to $\phi_\calBb{\in}\, q=\{0,1,2\}$ (i.e. we can have $0,1$ or $2$ excitations in $\calBb$):
	\begin{align}
		&E_0=\frac{1}{\sqrt{N{-}2}}\sum_{j=2}^{N{-}1}
		\begin{pmatrix}0&a_j^N\\0&0\end{pmatrix},\\
		&E_1=\frac{1}{\sqrt{N{-}2}}\sum_{j=2}^{N{-}1}
		\begin{pmatrix}b_{1j}^{kN}&0\\0&a_j^N\end{pmatrix},
		&\text{for}\;k&=\{1,\hdots, N{-}1\},\nonumber\\
		&E_2=\frac{1}{\sqrt{N{-}2}}\sum_{j=2}^{N{-}1} 
		\begin{pmatrix}0&0\\ b_{1j}^{kl}&0\end{pmatrix},
		&\text{for}\;k<l&=\{1,\hdots, N{-}1\}.\nonumber
	\end{align}
	where we have introduced the two site transition amplitudes $U_2^{ij}(t)=b_{i_1i_2}^{j_1j_2}(t)$ for all the pairs $\{i,j\}$.
	
	As in the previous example, the transfer fidelity can be written as a quadratic function of $\cos \theta$, and, again, inverting this relation and using \Cref{pdftransf}, it is possible to obtain an analytic expression of $\pdf(F)$, see \Cref{appA} for details. 
	In this case, also the average fidelity $\lavg F\ravg$ 
	depends on the $2$-excitations transition amplitudes:
	\begin{equation}
		\lavg F\ravg=\frac{1}{3}+\frac{1}{6 (N-2)}\sum _{k=1}^{N-1} \left|\sum _{j=2}^{N-1} a_j^k+b^{kN}_{1j}\right|^2.
	\end{equation}
	
\Cref{1_qubit_not_vacuum} shows the probability distribution corresponding to three protocols at fixed average fidelity $\langle F \rangle=0.99$. The shapes of the  distributions are different:  the $B$-protocol displays a minimum fidelity greater than the $W$ and $P$ protocols,  but, for most of the states, achieves a lower fidelity than the $W$- and $P$- ones, which, on the other hand, display a maximum achievable fidelity. 

	\subsection{Two-qubit transfer}\label{2Qubit}
	Consider now  transfer of a two-qubit quantum state: In this case the dynamics takes place in three $q$ subspaces and the Kraus operators act on a two-qubit subspace. Assuming $\ket{\psi_\calAb}$ to be initially in the vacuum state ($q_\calAb=0$), the three classes of Kraus operators are now given by:
    \begin{subequations}
	\begin{align}\label{kraus2}
		\centering
		&E_0=
		\begin{pmatrix}
			1 & 0 & 0 & 0 \\[0.5ex]
			0 & a_{N}^2 & a_{N}^1 & 0 \\[0.5ex]
			0 & a_{N{-}1}^2 & a_{N{-}1}^1 & 0 \\[0.5ex]
			0 & 0 & 0 & b_{N{-}1N}^{12} \\
		\end{pmatrix},
		\\
		&E_1^j=
		\begin{pmatrix}
			0 & a_{j}^2 & a_{j}^1 & 0 \\[0.5ex]
			0 & 0 & 0 & b_{jN}^{12} \\[0.5ex]
			0 & 0 & 0 & b_{jN{-}1}^{12} \\[0.5ex]
			0 & 0 & 0 & 0 \\
		\end{pmatrix},~j=\{1,\hdots,N{-}2\},
		\\
		&E_2^{kj}=
		\begin{pmatrix}
			0 & 0 & 0 & b_{kj}^{12} \\[0.5ex]
			0 & 0 & 0 & 0 \\[0.5ex]
			0 & 0 & 0 & 0 \\[0.5ex]
			0 & 0 & 0 & 0 \\
		\end{pmatrix},~k<j=\{1,\hdots,N{-}2\}.
	\end{align}
    \end{subequations}
	
	Writing a family of two-qubit states in the Schmidt form,
	\begin{equation}\label{param}
		\ket{\psi_\calA}=\sqrt{\frac{1-s}{2}}\ket{00}+\sqrt{\frac{1+s}{2}}\ket{11}, \quad -1 \leq s \leq 1
	\end{equation}
      we assure that
     	any other pure state can be obtained form it by local unitary transformations. Notice that the parameter $|s|$ quantifies the separability of the state, as the concurrence of $\ket{\psi_\calA}$
            reads $C=\sqrt{1-s^2}$.
	Due to \Cref{fid2} the fidelity now has the form
	\begin{equation}\label{fid4}
		F(t,U_1,U_2,C)=\sum_{\phi_\calBb}\vert\bra{\psi_\calA}U_1^\dagger U_2^\dagger \,E_{\phi_\calBb}(t)\,U_1 U_2\ket{\psi_\calA}\vert^2.
	\end{equation}
	Averaging over all possible local unitaries, the fidelity turns out to be a function of concurrence alone:
	\begin{equation}\label{fid_2Qubits}
		F(t,C)=\int\! \!F(t,u_1,u_2,C) du_1 du_2=A(t)-B(t)C^2
	\end{equation}
	where $A(t)$ and $B(t)$ are functions involving the transition amplitudes $a$s and $b$s entering the Kraus operators in \Cref{kraus2}. The lenghty expressions of these functions are reported in \Cref{app2qubit}. 
	From \Cref{fid_2Qubits} it is straightforward to see that the minimum fidelity (averaged over entanglement) achievable through the specific protocol is given by the minimum between $A(t)$ and $A(t)+B(t)$.

\begin{figure}[t]\centering
		\includegraphics[width=0.95\linewidth]{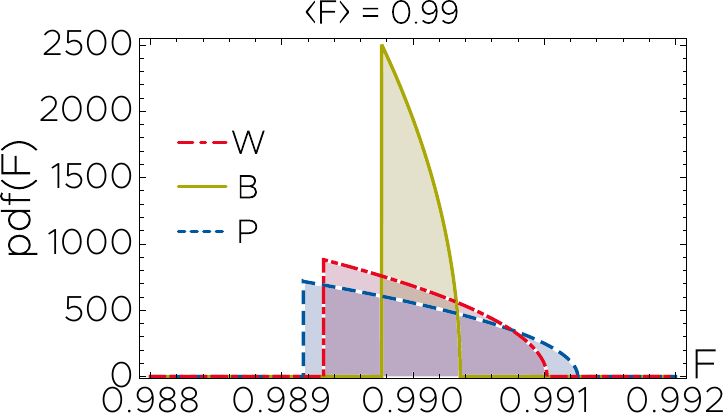}
		\caption{\label{pdf_of_F_2} Probability distribution of the fidelity for the 2-qubits transfer averaged over local unitary operations at fixed average fidelity $\langle F \rangle=0.99$. The minimum fidelity of the protocol $B$ is greater than this achieved in protocols $W$ and $P$. 
       Data are obtained for  $N=9$ and $h_0=200$ for protocol $B$ and $J_0=1/200$ for protocol $W$.}
	\end{figure}

	Making use of the distribution of  concurrence for two-qubit random pure states~\cite{ZS01,cappellini2006a}: 
	\begin{equation}\label{pdf_C}
		\pdf(C)=3C\sqrt{1-C^2}
	\end{equation}
	and inserting into \Cref{pdftransf}, we derive the distribution of fidelity for the transfer of a two-qubit state with any degree of entanglement,
	\begin{equation}
		\pdf(F)=\frac{3}{2|B(t)|}\sqrt{\frac{F-A(t)}{B(t)}}
		\label{PDF2qubits}
	\end{equation}
	with $(F{-}A(t))/B(t)>0$.\\
 Two-qubit fidelity distribution derived in \Cref{PDF2qubits} is used to obtain  \Cref{pdf_of_F_2} relative to the cases of the three different protocols ($W,B,P$) at fixed average fidelity $\langle F \rangle=0.99$.  The three distributions have the same shape, but that of the $B$ protocol is narrower, with a minimum fidelity greater than that obtained by the $W$ and $P$ protocols. 
	
	\section{Discussion and Conclusions}
 \label{conclusions}
	To characterize the quality of QST, previous works have considered the average transfer
    fidelity 
    between the initial state and its image with respect to a state transfer protocol, or, in a few cases, also the variance of fidelity
    ~\cite{longhi2019a,apollaro2022b}. In this contribution
    we go beyond these results and 
    focus on the entire probability distribution of fidelity for various state transfer protocols.  It is demonstrated that transmission protocols, which are equivalent with respect to the average transfer fidelity,  display different behaviors in terms of their reliability, and worst case performances.
	
     We exploited a global symmetry of the Hamiltonian describing the state transfer in order to evaluate analytically the probability distribution function of the fidelity for one and two qubits states. We derived the corresponding distributions of fidelity using explicit expressions for the Kraus operators describing the dynamical map that sends the input into the output state. Allowing for {\it non-optimal} scenarios, like errors in the reading times, we showed how different protocols, although resulting in the same average fidelity, can display very different distributions. Furthermore, for the one-qubit QST, we found an exact relation between the  sender-receiver transition amplitude and the minimum fidelity of the protocol.
     
	For the two-qubit transfer, we used the entanglement probability distribution of the input state to characterize the PDF of the entanglement averaged fidelity. We have left the examination of the mixed state transfer and the potential presence in the channel of noise and/or dissipation that may impact the dynamics for future analysis, as well as the extension of the transfer protocols for a larger number of subsystems,  $n>2$.
	
	Finally, considering the general framework presented for obtaining one- and two-qubit distribution of  fidelity, our results can be straightforwardly applied to state transfer protocols different that those considered here. 
	\vspace{6pt} 
	
	\acknowledgments 
	TJGA ackowledges funding by the European flagship on quantum technologies (‘ASPECTS’ consortium 101080167).
	MC acknowledges funding by the Tertiary Education Scholarships Scheme and the MCST Research Excellence Programme 2022 on the QVAQT project at the University of Malta.
	SL acknowledge support by MUR under PRIN Project No. 2017 SRN-BRK QUSHIP.
 KZ was supported by the QuantERA II Programme  
 financed by the National Science Centre, Poland, under the contract number 2021/03/Y/ST2/00193.
	The views and opinions expressed are however those of the authors only and do
	not necessarily reflect those of the European Union. Neither the European Union nor the granting authority can
	be held responsible for them.
	
	\appendix
	\section{Calculation of probability distribution of fidelity}\label{appA}
	In this section we provide the details for the evaluation of the PDF of the fidelity for the different cases analyzed in the main text.
	\subsection{\texorpdfstring{$\ket{\psi_\calAb}=\text{vacuum}$}{Vacuum} }
	Firstly note that \Cref{fid3} can be recast as a quadratic function of $\cos\theta$ 
	\begin{equation}
		\label{eq_fid_1}
		F=\frac{\vert a_1^N\vert^2{-}\text{Re}[a_1^N]}{2}\cos ^2(\theta ){+}
		\frac{1{-}\vert a_1^N\vert^2}{2} \cos (\theta )+\frac{1{+}\text{Re}[a_1^N]}{2},
	\end{equation} 
	which is easily solved for $\theta$.
	\\
	Defining $a{=}(|a_1^N|^2-\text{Re}[a_1^N])/2$, $b{=}(1-|a_1^N|^2)/2$  and $c{=}(1+\text{Re}[a_1^N])/2
	$ we find the expression of $\theta(F)\equiv F^{-1}$ to be
	\begin{equation}
		\theta^{(\pm)}(F)=-\frac{b}{2a}\pm \frac{\sqrt{\Delta(F)}}{2a}
	\end{equation}
	with $\Delta(F)=b^2-4a(c-F)$.\\
	Applying \Cref{pdftransf} we obtain
	\begin{equation}
		\pdf(F)=\pdf^{\,(-)}(F)+\pdf^{\,(+)}(F),
	\end{equation}
	where each term is defined for $\abs{\theta(F)^{(\pm)}}\leq 1$ and is given by
	\begin{equation*}
		\pdf^{\,(\pm
			)}(F)=\frac{1}{2}\sqrt{1{-}\theta^{(\pm)}(F)^2}\,\abs{\left(\Delta(F)\sqrt{1{-}\theta^{(\pm)}(F)^2}\right)^{-1}}.
	\end{equation*}
	\subsection{\texorpdfstring{$\ket{\psi_\calAb} \neq \text{vacuum}$}{Not Vacuum}}
	As mentioned in the main body of the paper, in this case the fidelity is also a quadratic function of $\cos\theta$,
	\begin{align}
		F(\theta)&=\left(\frac{1}{4}-\frac{3\alpha+\beta-3\gamma}{4(N-2)}\right)\cos^2 \theta \nonumber\\&+
		\left(\frac{1}{2}-\frac{\alpha+\gamma}{2(N-2)}\right)\cos \theta\nonumber\\
		&+
		\left(\frac{1}{4}+\frac{\alpha+\beta-\gamma}{4(N-2)}\right),
	\end{align}
	with $\alpha=\abs{\sum_{j=2}^{N-1}a_j^N}^2$, $\beta=\sum_{k=1}^{N-1}\abs{\sum_{j=2}^{N-1}a_j^N+b_{1j}^{kN}}^2$ and $\gamma=\sum_{k=1}^{N-1}\abs{\sum_{j=2}^{N-1}b_{1j}^{kN}}^2$.
	
	Redefining the coefficients $a,b,c$ it is possible to proceed as in the previous section in the evaluation of the $\pdf(F)$.

\subsection{Two-qubit transfer}\label{app2qubit}
As discussed in the main text, averaging over local unitaries acting on the two qubits input state, the fidelity can be written as a function of the initial concurrence $C$ (see \Cref{fid_2Qubits}) as:
\begin{equation}
		F(t,C)=A(t)-B(t)C^2
\end{equation}
where the time dependent coefficients $A$ and $B$ are given by
\begin{eqnarray}
A(t){=}
&&\frac{1}{72}\big(14+2\sum_{j=1}^{N-2}|a^{1}_{j}+b^{12}_{jN}|^2+|a^{2}_{j}+b^{12}_{jN-1}|^2 +\nonumber\\
&&2\,\text{Re}\big [2a^{1}_{N-1}a^{*2}_{N}{+}2b^{12}_{N-1N}{+}4(a^{1}_{N-1}{+}a^{2}_{N})(1{+}b^{*12}_{N-1N})\big ]+\nonumber\\
&&6\big(|a^{1}_{N-1}|^2+|a^{2}_{N}|^2+|b^{12}_{N-1N}|^2\big)  \big)
\label{A}\end{eqnarray}
\begin{eqnarray}
B(t){=}
&&\frac{1}{72}\big({-}10 + 5\sum_{j=1}^{N-2}|a^{1}_{j}+b^{12}_{jN}|^2+|a^{2}_{j}+b^{12}_{jN-1}|^2-\nonumber\\
&&2\,\text{Re}\big[4a^{1}_{N-1}a^{*2}_{N}{+}4b^{12}_{N-1N}{-}(a^{1}_{N-1}{+}a^{2}_{N})(1{+}b^{*12}_{N-1N})\big]+\nonumber\\
&&6(|a^{1}_{N-1}|^2+|a^{2}_{N}|^2+|b^{12}_{N-1N}|^2)\big)
\label{B}\end{eqnarray}
For example, in the three instances reported in \Cref{PDF2qubits}, respectively for the three protocols $W,B,P$, at fixed average fidelity $\langle F \rangle=0.99$, the values of these functions are:
		\begin{equation}
			\begin{cases}
				B)\qquad &$A=0.9904$ \, ,\, $B=-0.0006$\\
				W)\qquad &$A=0.9912$ \, ,\, $B=-0.0021$\\
				P)\qquad &$A=0.9910$ \, ,\, $B=-0.0017$.
			\end{cases}
		\end{equation}

	\section{Details of the B-, W- and P-protocols}\label{appB}

 Recalling equation \Cref{BoseH} of the main text, 
	the three protocols  $(W,B,P)$ are defined by different choices of the couplings and the magnetic fields (see \Cref{chains_sketch}) as reported in \Cref{table}. For all the three cases one has $\Delta_{ij}=0$. The weak-coupling protocol $W$ is defined by having the first and last couplings much smaller than the rest ($J_0\ll J$) and no local magnetic fields.

 \begin{table}[h!]
		\resizebox{0.4\textwidth}{!}
		{\begin{tabular}{c||c|c}
				protocol           & couplings                 & magnetic field       \\
				\hline\hline
				\multirow{2}{*}{W} & \multicolumn{1}{c|}{$J\delta_{i,i{+}1}$}     & \multirow{2}{*}{$B_i{=}0$} \\
				& \multicolumn{1}{c|}{$J_{1,2}{=}J_{N{-}1,N}=J_0$} &                      \\
				\hline
				\multirow{2}{*}{B} & \multirow{2}{*}{$J\delta_{i,i{+}1}$}     & \multicolumn{1}{c}{$B_i{=}0$} \\
				&                                          & \multicolumn{1}{c}{$B_2{=}B_{N{-}1}=h_0$}                    \\
				\hline
				\multirow{2}{*}{P} & \multirow{2}{*}{$J\sqrt{i(N-i)}\delta_{i,i{+}1}$}    & \multirow{2}{*}{$B_i{=}0$} \\
				&                           &                     
		\end{tabular}}
		\caption{Values of coupling constants and local magnetic fields that define the three protocols considered in this work.}\label{table}
	\end{table}

\begin{figure}[t]\centering
		\includegraphics[width=1.0\linewidth]{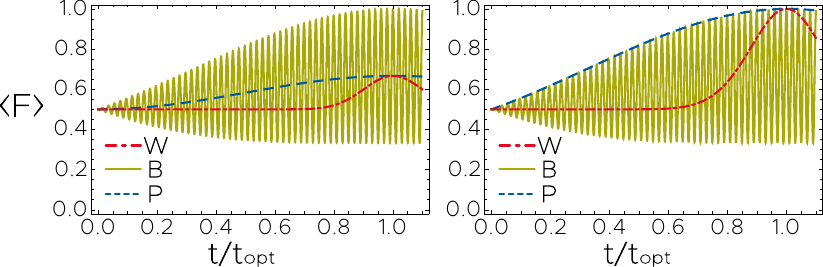}
		\caption{ Average fidelity relative to three different protocols
        plotted against a rescaled time with respect to the relative optimal time $t_{\text{opt}}$. 
  No auxiliary magnetic field is applied on the chain (left panel) compared to a constant magnetic field assuring  that at the optimal time $t_\text{opt}$ the average fidelity is as close as possible to unity (right). No significant differences occur for protocol $B$, whereas for  protocols $W$ and $P$  the auxiliary field is necessary. Both panels refer to a chain of $N=22$ sites with $h_0=200$ for protocol $B$ and $J_0=1/200$ for protocol $W$.  }\label{average}
	\end{figure}

The barrier protocol $B$ has uniform couplings in all the chain and strong magnetic fields only on the second and the last but one sites ($h_0\gg J$). The perfect protocol $P$ is characterized by engineered couplings and no magnetic fields.

	For the one qubit QST with the chain initially in the vacuum, as in \Cref{1qubit_vacuum}, expressing the transition amplitude in polar form $a_1^N=r e^{i \phi}$, the fidelity averaged over all possible pure state is given by \cite{bose2003a}:
	\begin{equation}
		\lavg F \ravg=\frac{1}{2}+\frac{r\cos{\phi}}{3}+\frac{r^2}{6}
	\end{equation}
	The phase $\phi$ of the transition amplitude $a_1^N$ can be adjusted applying a uniform magnetic field such that at the optimal time $t_{\text{opt}}$ one retrieves $\cos{\phi}=1$. This is shown in \Cref{average}: on the left we report the behavior of $\lavg F \ravg$ for the three protocols without the auxiliary magnetic field, and on the right an uniform magnetic field $B_{\text{aux}}=\text{Arg}(a_1^N(t_{\text{opt}}))/t_\text{opt}$ is applied.
 
        \bibliographystyle{apsrev4-2}
	\bibliography{pdfofF}
	
\end{document}